\documentclass[amsmath,amssymb,showpacs,twocolumn]{revtex4}
\usepackage{epsfig}
\usepackage{graphicx}
\usepackage{color}

\begin{document}

\title{Dynamics of the sub-Ohmic spin-boson model: a time-dependent variational study}

\author{Ning Wu$^1$, Liwei Duan$^1$, Xin Li $^{1,2}$, Yang Zhao$^{1}$\footnote{Electronic address:~\url{YZhao@ntu.edu.sg}}}
\date{\today}

\address{$^{1}$School of Materials Science and Engineering, Nanyang Technological University, Singapore 639798, Singapore\\
$^2$Institute of High Energy Physics,
Theoretical Physics Center for Science Facilities,\\
Chinese Academy of Sciences, 100049 Beijing, China}

\begin{abstract}
The Dirac-Frenkel time-dependent variation is employed to probe the dynamics of the zero temperature sub-Ohmic spin-boson model with strong friction utilizing the Davydov ${\rm D}_1$ ansatz. It is shown that initial conditions of the phonon bath have considerable influence on the dynamics. Counterintuitively, even in the very strong coupling regime, quantum coherence features still manage to survive under the polarized bath initial condition, while such features are absent under the factorized bath initial condition.
In addition, a coherent-incoherent transition is found at a critical coupling strength
$\alpha\approx0.1$ for $s=0.25$ under the factorized bath initial condition. We quantify how faithfully our ansatz follows the Schr\"{o}dinger equation, finding that the time-dependent variational approach is robust for strong dissipation and deep sub-Ohmic baths ($s\ll1$).
\end{abstract}

\maketitle
\section{Introduction}
It is of fundamental importance to study macroscopic behavior of open quantum systems under the influence of dissipative baths that they are inevitably in contact with \cite{Leggett, Weiss}. As an open quantum system and its thermal bath with a finite number of modes together form an isolated system which conserves the energy, according to the Poincar\'{e} recurrence theorem \cite{Bocchieri}, the quantum system will eventually return to a state very close to its initial state. To circumvent this recurrence, the bath has to be expanded to contain an infinite number of phonon modes.  The simplest model of a quantum system is a two-level system $\hat{H}_{\rm S}=-\Delta\sigma_x/2$, and Caldeira and Leggett\cite{Cal-Leg1} have shown that a bath of harmonic oscillator, $\hat{H}_{\rm B} =\sum_l \omega_l b_l^\dag b_l$, provides a very good approximation to real dissipative systems. Here $\sigma_x$ is a Pauli matrix, and $b_l^\dag$ ($b_l$) is the boson creation (annihilation) operator. In general, it is assumed that the two-level system is linearly coupled to the bath, $\hat{H}_{\rm SB}=-\sigma_z\sum_l \lambda_l(b^\dag_l+b_l)/2$.
In the absence of the bath, the quantum two-level system will oscillate between the states $|+\rangle$ and $|-\rangle$ with frequency $\Delta$ ($|\pm\rangle$ are the two eigenstates of $\sigma_z$), a quantum phenomenon which has no classical analog.
Putting together all three Hamiltonian terms,
\begin{equation}\label{hami}
\hat{H}=-\frac{\Delta}{2}\sigma_x+\sum_l \omega_l b_l^\dag b_l+\frac{\sigma_z}{2}\sum_l \lambda_l(b^\dag_l+b_l)~.
\end{equation}
The combined Hamiltonian described by Eq.~(\ref{hami}) is called the spin-boson model.

Despite its simplicity, the spin-boson model has been widely discussed in condensed phase physics and chemistry, ranging from the process of electron transfer \cite{Marcus} to quantum entanglement \cite{Costi} between a qubit \cite{Khveshchenko} and its bath. The coupling between the two-level system and the harmonic bath is completely specified by the spectral function
\begin{equation}\label{spectra}
J(\omega)=\sum_l \lambda^2_l \delta(\omega-\omega_l)=2\alpha\omega_c^{1-s}\omega^s e^{-\omega/\omega_c},
\end{equation}
with a cutoff frequency $\omega_c$ and a dimensionless constant $\alpha$ measuring the strength of the coupling. A bath described by a spectral function with $s=1$ is referred to as an Ohmic bath. Both static and dynamical properties of the spin-boson model with an Ohmic bath are well understood \cite{Leggett,Weiss}. It is found that there exists a quantum phase transition from a non-degenerate delocalized phase to a doubly degenerate localized phase and a turnover from a coherent phase to an incoherent one. The critical coupling for the quantum phase transition is $\alpha_c\approx1$, and the critical coupling for the turnover to occur is $\alpha_{\rm CI}\approx0.5$.
In the Ohmic regime, the spin-boson model can be readily mapped onto the anisotropic Kondo model using bosonization techniques\cite{Leggett}, and known results can be borrowed from the Kondo model. However, success still eludes us in arriving at a correct description of the spin-boson physics in the sub-Ohmic regime ($0<s<1$).

Several sophisticated numerical methods have been used to study the sub-Ohmic spin-boson model. An incomplete list includes: The numerical renormalization group method developed by Wilson \cite{Wilson} which reveals a continuous quantum phase transition for all $0<s<1$ and weakly damped coherent oscillations on short time scales in the localized phase \cite{Bulla}; numerically exact real-time path integral method with quasiadiabatic propagator revealing effective dynamic asymmetry in the presence of a sub-Ohmic bath \cite{Makri,Nalbach}; quantum Monte Carlo method which determines the critical exponents for $s<1/2$ \cite{Winter}; the sparse polynomial space representation method which is based on the exact diagonalization to obtain numerical results\cite{Alvermann}; the real-time path integral Monte Carlo techniques which show that the coherent phase exists even in strong dissipation for $s<1/2$ \cite{Egger,Kast}; the numerically exact multilayer multiconfiguration time-dependent Hartree method (ML-MCTDH) which shows that the transition of the dynamics from weakly damped
coherent motion to localization upon increase of the system-bath coupling strength \cite{Wang1,Wang2}.

Unlike in the Ohmic regime, dynamics of sub-Ohmic spin-boson model is very sensitive to the initial conditions. There are two initial conditions of interest: one is the factorized initial condition with the bath in its vacuum state initially; the other is the polarized initial condition consistent with typical experimental scenarios \cite{Weiss}, under which the system is prepared in the ground state of $H_{\rm B}+H_{\rm SB}|_{\sigma_z=1}$.
Many recent studies use the polarized initial condition \cite{Winter,Alvermann,Nalbach,Kast}. The typical time scale of the spin dynamics is $~1/\Delta$,  the time it takes to hop from a spin state to another. On the other hand, the reorganization energy \cite{Lucke}, which describes the change in population disparity between the two states $|+\rangle$ and $|-\rangle$ as one goes from one initial condition to the other, is given as $\int_0^\infty d\omega {J(\omega)}/{\omega}=2\alpha\omega_c\Gamma(s)$, where $\Gamma(s)$ is the gamma function of $s$. Ref.~\cite{Lucke} shows that the difference in spin dynamics under the two initial conditions is negligible for the Ohmic bath in the scaling limit $\Delta\ll\omega_c$. However, reorganization energies in the sub-Ohmic regime are larger than those in the Ohmic regime. It implies that much more time is needed for a sub-Ohmic bath to return to thermal equilibrium, and therefore the spin dynamics corresponding to the two initial conditions will display sizeable differences for certain parameter space (${\Delta}/{\omega_c},s$). Ref.~\cite{Nalbach} has confirmed such a physical picture. It is commonly accepted that the increasing of the spin-bath coupling will eventually turn quantum coherent oscillations into classical-like damping, a picture supported by the aforementioned numerical approaches for $1/2<s<1$. However, Ref.~\cite{Kast} has recently claimed that such a picture may not always hold. Numerical data in the strong coupling regime, for example, with $\alpha=30\alpha_c$, show that the coherent phase exists for exponents up to $s=0.49$. Such an ``anomalous" result warrants further investigations.

In this work, we adopt a variational framework to study the zero temperature dynamical properties of the sub-Ohmic spin-boson model. This is motivated by two facts: (i) The spin-bath interactions are formally identical to the exciton-phonon coupling in a quasiparticle named a polaron, which is generally described by the Holstein model \cite{Holstein}
\begin{eqnarray}
\hat{H}_{\rm Holstein}&=&\sum_m\epsilon_ma^\dag_ma_m+\sum_{m\neq n}J_{mn}(a^\dag_ma_n+a^\dag_na_m)\nonumber\\
&&+\sum_{qm}g_qa^\dag_ma_m(b_q+b^\dag_q)+\sum_q\omega_qb^\dag_qb_q
\end{eqnarray}
where $a^\dag_m$ ($a_m$) is the exciton creation (annihilation) operator, $J_{mn}$ is the hopping integral, and $g_q$ labels the exciton-phonon coupling strength; (ii) A time-dependent variational approach based on the Davydov ans\"{a}tze has been widely used for describing the dynamics of Holstein systems. As a semi-classical approach for studying energy transport in deformable molecular chains, those ans\"{a}tze \cite{Scott} were put forward by Davydov and coworkers as two trial wave functions, namely, the Davydov ${\rm D}_1$ and ${\rm D}_2$ ans\"{a}tze. The first of Davydov's ans\"{a}tze has the form
\begin{equation}
|{\rm D}_1(t)\rangle=\sum_n\alpha_n(t)a_n^\dag|0\rangle_{\rm ex}\exp[\sum_q(\lambda_{nq}(t)b^\dag_q-h.c.)]|0\rangle_{\rm ph},
\end{equation}
where $\alpha_n(t)$ and $\lambda_{nq}(t)$ are the variational parameters representing the exciton amplitude and the phonon displacements at the $n$th site, respectively, and $|0\rangle_{\rm ex}$ and $|0\rangle_{\rm ph}$ denote the exciton and the phonon vacuum states, respectively.
The second of Davydov's ans\"{a}tze is given by
\begin{equation}
|{\rm D}_2(t)\rangle=\sum_n\alpha_n(t)a_n^\dag|0\rangle_{\rm ex}\exp[\sum_q(\beta_{q}(t)b^\dag_q-h.c.)]|0\rangle_{\rm ph},
\end{equation}
Note that the phonon-displacement parameter in the much simplified $|D_2(t)\rangle$, $\beta_{q}(t)$, is independent of the site index $n$. The Davydov ans\"{a}tze and their variants have also been applied successfully to study the one-dimensional Holstein polaron by Zhao and coworkers \cite{Zhao,Sun,pssc,Ye}. To probe polaron dynamics of the Holstein system using Davydov's ans\"{a}tze \cite{Sun,pssc}, we have employed the Dirac-Frenkel time-dependent variational principle, a powerful technique to obtain approximate dynamics for quantum systems for which exact solutions are elusive \cite{Dirac}.

By exploiting the analogy between the spin-boson model and the Holstein molecular crystal model, we propose a time-dependent trial wave function very similar to the Davydov ${\rm D}_1$ ansatz, and seek to develop an accurate description for dynamical properties of the sub-Ohmic spin-boson model with $0<s<1/2$ under both the polarized and factorized initial conditions.
It was pointed earlier \cite{satoshi} that the Davydov ans\"{a}tze bear close resemblance to
a multiconfigurational ansatz which contains more than one Slater determinant.
We note that the ansatz employed in this work shares many characteristics with the multi-configurational Gaussian wave packets \cite{Burghardt-1999,Burghardt-2003,Martinazzo-2006,Sawada-1986,Shalashilin-2008}, used in a variant of the powerful multiconfiguration time-dependent Hartree technique (MCTDH) \cite{MCTDH}, also known as the G-MCTDH method.
Proposed in Ref.~\cite{Burghardt-1999} and developed further in Refs.~\cite{Burghardt-2003,Burghardt-CPL},
the G-MCTDH method extends MCTDH to higher-dimensional systems by including a moving basis of Gaussian functions while restricted to an optimally chosen subset of the bath modes. It has been successfully applied to describe dynamics of the Henon-Heiles potential model \cite{Burghardt-CPL} and the oscillator-bath model \cite{Burghardt-2003}.

In the Holstein molecular crystal model, a similar reorganization energy \cite{Sun} can be calculated from the phonon spectral density function using $\int^\infty_0d\omega {J(\omega)}/{\omega}$, and is known to be proportional to the Huang-Rhys factor \cite{Huang}, which measures the exciton-phonon coupling strength. For the factorized initial conditions, our previous studies on the Holstein model reveal that the Davydov ans\"{a}tz are especially accurate in the strong exciton-phonon coupling regime\cite{Zhao}, a fact that will also be confirmed by calculating the relative deviation for the sub-Ohmic spin-boson model studied here \cite{Sun}. Furthermore, for a given spin-bath coupling strength, a smaller $s$ yields a smaller relative deviation,
inferring the highest accuracy of our ansatz in the deep sub-Ohmic regime $s\ll 1$.
A similar trend is believed for the polarized initial conditions.

The paper is organized as follows. In Sec.~II, we propose a Davydov-like ansatz, and use the Dirac-Frenkel variational method to obtain the equations of motion for its parameters. In Sec.~III, we present the numerical results, which show the quantum coherence and entanglement for spin-boson model. The effect of different initial conditions is discussed in this section. In Sec.~IV, conclusions are drawn.

\section{Methodology}
Note that the spin-boson model can be viewed as a two-site Holstein model with an infinite number of phonon modes in the one-exciton subspace.
This equivalence between the exciton-phonon coupling and the spin-boson interaction naturally leads to
a trial wave function similar to Davydov ${\rm D}_1$ ansatz for the spin-boson model
\begin{eqnarray}
|D_s(t)\rangle=&&A(t)|+\rangle\exp[\sum_l (f_l(t)b_l^\dag-h.c.)]|0\rangle_{\rm ph}\nonumber\\
\label{trial func}
&+&B(t)|-\rangle\exp[\sum_l (g_l(t)b_l^\dag-h.c.)]|0\rangle_{\rm ph},
\end{eqnarray}
where $A(t)$ and $B(t)$ are complex variational parameters representing occupation amplitudes in states $|+\rangle$ and $|-\rangle$, respectively, and $f_l(t)$ and $g_l(t)$ label the corresponding complex phonon displacements of the l-th phonon mode.
In this work, we choose the Lagrangian formalism of the Dirac-Frenkel variational principle to obtain equations of motion for the variational parameters. The Lagrangian associated with the trial state $|D_s(t)\rangle$ is given as
\begin{equation}\label{lag}
L=\langle D_s(t)|\frac{i}{2}\frac{\overleftrightarrow{\partial}}{\partial t}-\hat{H}|D_s(t)\rangle~.
\end{equation}
Substituting the trial state $|D_s(t)\rangle$ into the Lagrangian (\ref{lag}), we arrive at the Lagrangian for the spin-boson model
\begin{eqnarray}
L=&&\frac{i}{2}(A^*\dot{A}-\dot{A}^*A+B^*\dot{B}-\dot{B}^*B)\nonumber\\
&+&\frac{i}{2}\sum_l[|A|^2(f_l^*\dot{f_l}-\dot{f_l}^*f_l)+|B|^2(g_l^*\dot{g_l}-\dot{g_l}^*g_l)]\nonumber\\
&-&\langle D_s(t)|\hat{H}|D_s(t)\rangle,
\end{eqnarray}
where
\begin{eqnarray}
&&\langle D_s(t)|\hat{H}|D_s(t)\rangle=\sum_l\omega_l(|A|^2|f_l|^2+|B|^2|g_l|^2)\nonumber\\
&-&\frac{\Delta}{2}(A^*Be^{\sum_l f_l^*g_l}+AB^*e^{\sum_lf_lg_l^*})e^{-\frac{1}{2}\sum_l(|f_l|^2+|g_l|^2)}\nonumber\\
&+&\frac{1}{2}\sum_l\lambda_l[|A|^2(f_l+f_l^*)-|B|^2(g_l+g_l^*)]~.
\end{eqnarray}
The Dirac-Frenkel time-dependent variational principle yields the equations of motion for $A,B,f_l$ and $g_l$
\begin{equation}\label{lag eq}
\frac{d}{dt}\left(\frac{\partial L}{\partial \dot{u}_n^*}\right)-\frac{\partial L}{\partial u_n^*}=0,
\end{equation}
where $u_n^*$ denotes the complex conjugate of variational parameters $u_n$, which can be $A,B,f_l$ or $g_l$.
From Eq.~(\ref{lag eq}),
one arrives at the equations of motion for $A(t)$ and $B(t)$
\begin{eqnarray}
0=&&i\dot{A}+i\frac{A}{2}\sum_l(\dot{f}_lf_l^*-\dot{f}_l^*f_l)-\frac{A}{2}\sum_l\lambda_l (f_l+f_l^*)\nonumber\\
\label{eq A}
&+&\frac{\Delta B}{2}e^{\sum_l \left[f_l^*g_l-\frac{1}{2}(|f_l|^2+|g_l|^2)\right]}-A\sum_l\omega_l|f_l|^2~,\\
0=&&i\dot{B}+i\frac{B}{2}\sum_l(\dot{g}_lg_l^*-\dot{g}_l^*g_l)+\frac{B}{2}\sum_l\lambda_l (g_l+g_l^*)\nonumber\\
\label{eq B}
&+&\frac{\Delta A}{2}e^{\sum_l \left[g_l^*f_l-\frac{1}{2}(|f_l|^2+|g_l|^2)\right]}-B\sum_l\omega_l|g_l|^2~.
\end{eqnarray}
Similarly, the equations of motion for $f_l(t)$ and $g_l(t)$ are given as
\begin{eqnarray}
&&iA\dot{f}_l-A\frac{\lambda_l}{2}-A\omega_l f_l\nonumber\\
\label{eq f}
&=&\frac{\Delta B}{2}(f_l-g_l)e^{\sum_l \left[f_l^*g_l-\frac{1}{2}(|f_l|^2+|g_l|^2)\right]}~,\\
&&iB\dot{g}_l+B\frac{\lambda_l}{2}-B\omega_l g_l\nonumber\\
\label{eq g}
&=&\frac{\Delta A}{2}(g_l-f_l)e^{\sum_l \left[g_l^*f_l-\frac{1}{2}(|f_l|^2+|g_l|^2)\right]}~.
\end{eqnarray}
Eqs.~(\ref{eq A}) and (\ref{eq B}) have been made use of to deduce the equations of motion for $f_l(t)$ and $g_l(t)$.
It is found from the equations (\ref{eq A}) and (\ref{eq B}) that
\begin{equation}
\frac{d}{dt}(|A|^2+|B|^2)=\frac{d}{dt}\langle D_s(t)|D_s(t)\rangle=0~.
\end{equation}
That the sum of $|A|^2$ and $|B|^2$ is conserved follows from early assignments of $A(t)$ and $B(t)$ in Eq.~(\ref{trial func}). Therefore, $|A|^2+|B|^2$, which is the norm of $|D_s(t)\rangle$, can be set to unity
\begin{equation}
|A|^2+|B|^2=1~.
\end{equation}
The equations of motion (\ref{eq A}--\ref{eq g}) give a complete description of the time evolution of $|D_s(t)\rangle$,
and therefore, the dynamics of the spin-boson model. In the spin-boson model, physical observables of interest are
\begin{equation}\label{P}
P_i(t)\equiv\langle\sigma_i\rangle=\langle D_s(t)|\sigma_i|D_s(t)\rangle~,~i=x,y,z.
\end{equation}
Here $P_x(t)$ describes the coherence between the $|+\rangle$ and $|-\rangle$ states, and $P_z(t)$, the population difference.
Upon substitution of the trial wave function (\ref{trial func}) into Eq.~(\ref{P}), we obtain
\begin{eqnarray}\label{pz}
P_z(t)=&&|A|^2-|B|^2,\\
\label{px}
P_x(t)=&&A^*Be^{\sum_l \left[f_l^*g_l-\frac{1}{2}(|f_l|^2+|g_l|^2)\right]}\nonumber\\
&+&AB^*e^{\sum_l \left[g_l^*f_l-\frac{1}{2}(|f_l|^2+|g_l|^2)\right]},\\
P_y(t)=&-&iA^*Be^{\sum_l \left[f_l^*g_l-\frac{1}{2}(|f_l|^2+|g_l|^2)\right]}\nonumber\\
&+&iAB^*e^{\sum_l \left[g_l^*f_l-\frac{1}{2}(|f_l|^2+|g_l|^2)\right].}\label{py}
\end{eqnarray}
Due to the invariance of $H$ under $\sigma_y\rightarrow-\sigma_y$, one usually has $P_y=0$ for the ground state or thermal averages. However, the time-dependent observable $P_y(t)$ is in general nonzero.

As mentioned in the Introduction, the initial condition has a vital influence on the dynamics of the spin-boson model with the sub-Ohmic bath. We assume that the spin is prepared in state $|+\rangle$ at $t=0$, or $A(0)=1$ and $B(0)=0$.
 At zero temperature, the factorized bath initial condition corresponds to a phonon vacuum state with $f_l(0)=g_l(0)=0$, while
 the polarized bath initial condition is one in which the bath phonons are in a displaced-oscillator state with $f_l(0)=g_l(0)=-{\lambda_l}/{2\omega_l}$.

When the phonon bath is absent, the dynamics of $P_z(t)$ will be fully coherent and has no classical component. The bath will induce decoherence, and for a quantum dissipative system such as the spin-boson model, its population difference has the form that $P_z(t)\sim\cos(\Omega t)e^{-\gamma t}$ \cite{Leggett,Weiss} on certain time scales. The oscillation $\cos(\Omega t)$ represents the quantum coherence. The exponential decay is classical friction effect that induced by the bath. The dynamics is said to be coherent if $\Omega\neq0$, otherwise it is incoherent. On the other hand, the steady state is said to be localized if $P_z(t\rightarrow\infty)\neq0$, otherwise it is delocalized $P_z(t\rightarrow\infty)=0$\cite{Bulla,Winter,Kast,Wu,Zhang}.

Another physical quantity of interest is the entanglement between the spin and the surrounding bath described by the von Neumann entropy $S$, also known as the entanglement entropy \cite{Bennett,zz}. In the spin-boson model, it is given as \cite{Costi,Amico}
\begin{equation}\label{entropy}
S=-\omega_+\ln\omega_+-\omega_-\ln\omega_-~,
\end{equation}
where
\begin{eqnarray}
\omega_\pm&=&\left(1\pm\sqrt{P_x^2+P_y^2+P_z^2}\right)/2\nonumber\\
\label{entropy term}
&=&\frac{1}{2}\pm\frac{1}{2}\sqrt{1+4|A|^2 |B|^2(e^{-\sum_l|f_l-g_l|^2}-1)}.
\end{eqnarray}
From Eq.~(\ref{py}),
it is clear that $P_y(t)=0$ if and only if $A^*Be^{\sum_l f_l^*g_l}$ is a real number, a condition that is satisfied in the ground state (or ensemble averages)  due to Hamiltonian invariance the transformation $\sigma_y\rightarrow-\sigma_y$ \cite{Costi}.

\section{Numerical results}

The spectral function (\ref{spectra}) gives full information for the spin-bath coupling $\lambda_l$. Together with equations of motion (\ref{eq A}--\ref{eq g}), the dynamics of spin-boson model could be deduced for given specific initial conditions. The $|+\rangle$ state is usually populated at $t=0$, i.e. $A(0)=1$ and $B(0)=0$. The initial conditions of the phonon bath are $f_l(0)=g_l(0)=0$ and $f_l(0)=g_l(0)=-{\lambda_l}/{2\omega_l}$ for factorized and polarized initial conditions, respectively. We have to solve the $2N_b+2$ equations of motion Eqs. (\ref{eq A}--\ref{eq g}), together with the $2N_b+2$ initial conditions mentioned above, where $N_b$ is the number of the phonon modes considered.

\begin{figure}[tbh]
\includegraphics[trim=0 5 0 0,clip=true,scale=0.35]{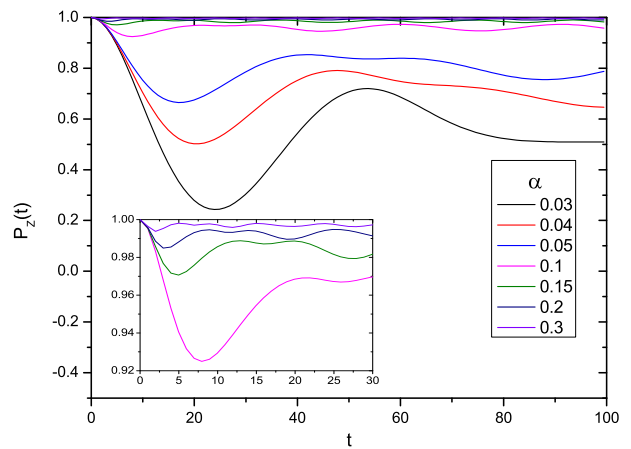}
\caption{Under the polarized bath initial conditions, the time dependent population difference $P_z(t)$ for $s=0.25$ ($\alpha_c \approx 0.022$) and $\Delta/\omega_c=0.1$ is presented. Seven values of $\alpha$ are taken. The inset is a magnified figure for large couplings. }
\end{figure}

We will adopt the homogeneous discretization procedure used in Ref.~\cite{Stock}. The frequencies of the $N_b$ harmonic modes are equally distributed in the frequency range $\omega\in[\omega_{\min},\omega_{\max}]$ with spacing $\Delta\omega=\omega_{\max}/N_b$ so that $\omega_{\min}=\omega_1=\Delta\omega$ and  $\omega_l=l\Delta\omega$.
The frequency spacing $\Delta\omega$ determines the Poincare recurrence time $T_{\rm p}= {2\pi}/{\Delta\omega}$ which must be greater than any time scale of interest \cite{poincare1,poincare2}. Throughout this work, we will use $N_b=20000$ and $\omega_{\max}=4\omega_c$, resulting in a recurrence time $T_{\rm p}=10000\pi$ that places our simulations at a safe distance from the Poincare recurrence.
Correspondingly, from the integration of the spectral density over $\omega$
\begin{eqnarray}
 \sum^{N_b}_{l=1}\lambda^2_l=\int^\infty_0d\omega J(\omega)\approx \sum^{N_b}_{l=1}J(\omega_l)\Delta\omega,
\end{eqnarray}
we obtain that $\lambda^2_l =J(\omega_l)\Delta\omega$.
It is found that under the factorized bath initial condition the simulation results are insensitive to the number of phonon modes.
Under the polarized initial condition, the number of the phonon modes has a considerable influence on the dynamics, which is especially true in the strong coupling regime. However, our numerical tests show that good convergence is reached when $N_b=20000$ for the time periods considered.

\begin{figure}[t]
\includegraphics[scale=0.55]{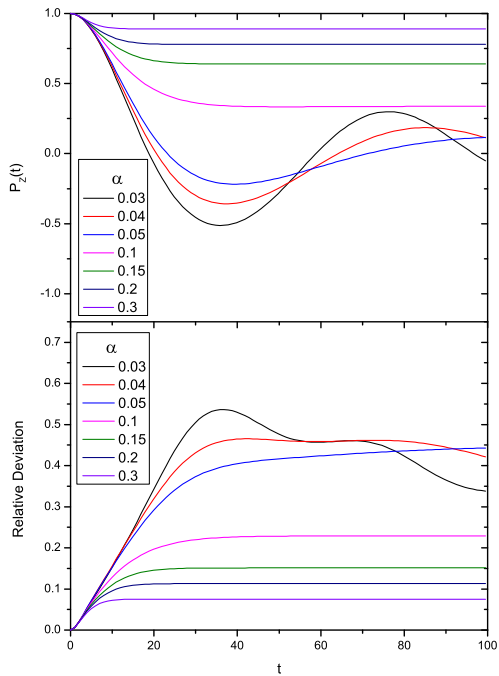}
\caption{Time dependent population difference $P_z(t)$ (upper panel) and relative deviation $\sigma (t)$ (lower panel) for $s=0.25$ ($\alpha_c \approx 0.022$) and $\Delta/\omega_c=0.1$ under the factorized bath initial conditions. Seven values of $\alpha$ are used for comparison.}
\end{figure}

As shown in Fig.~1, under the polarized initial condition, the population difference as a function of time, $P_z(t)$, manifests coherent oscillations even for very large couplings $\alpha=0.3\approx13\alpha_c$, where $\alpha_c\approx0.022$ \cite{Chin,Nalbach} is the critical coupling for the quantum phase transition. It is clearly seen that oscillatory behavior emerges even for very strong coupling far beyond $\alpha_c$. Furthermore, the characteristic oscillation frequency of $P_z(t)$ increases with increasing $\alpha$. Our results agree with those of Kast {\it et al.}~obtained using the real-time path integral Monte Carlo simulation\cite{Kast}. It is widely accepted that under the polarized initial condition, a quantum dissipative system such as the spin-boson model is expected to display classical over-damped behavior (or incoherent phase) at strong spin-bath couplings. Our results reveal that this is not the case for $0<s<0.5$. However, such apparent contradictions only appear for the polarized bath initial condition.
For the factorized initial bath condition, the persistent coherence does not occur.
The upper panel of Fig.~2 shows the population difference as a function of time, $P_z(t)$, under the factorized bath initial condition.
It is found that the critical coupling strength for the coherent-incoherent transition is $\alpha^{(f)}_{\rm CI}\approx0.1\approx4.5\alpha_c$.
In the lower panel of Fig.~2, we also plot the relative deviation \cite{Sun} of the trial state $|D_s(t)\rangle$ defined as
\begin{eqnarray}
\sigma (t) =\frac{\sqrt{\langle\delta(t)|\delta(t)\rangle}}{\bar{E}_{\rm bath}}
\end{eqnarray}
where $\bar{E}_{\rm bath}$ denotes
the average energy of bath within the time interval considered, and $|\delta(t)\rangle$ is the deviation vector quantifying how faithfully $|D_s(t)\rangle$ follows
the Schr\"{o}dinger equation:
\begin{eqnarray}
|\delta(t)\rangle=(i\partial_t-\hat{H})|D_s(t)\rangle.
\end{eqnarray}
In another word, the smaller the relative deviation $\sigma (t)$, the closer the trial state $|D_s(t)\rangle$ obeys
the Schr\"{o}dinger equation.
To compare with the ML-MCTDH method, a calculation is carried out for four values of $s$. The upper panel of Fig.~3 shows the time-dependent population difference $P_z(t)$ for $\alpha=0.2$, $\omega_c/\Delta=5$, in good agreement with corresponding results in Ref.~\cite{Wang2}.
As shown in the lower panel of Fig.~3, the relative deviation $\sigma (t)$ gradually increases with time before reaching a saturation value that have a strong dependence on the exponent $s$, indicating that the smaller the exponent $s$ is, the more accurate our ansatz becomes.

\begin{figure}[b]
\includegraphics[scale=0.55]{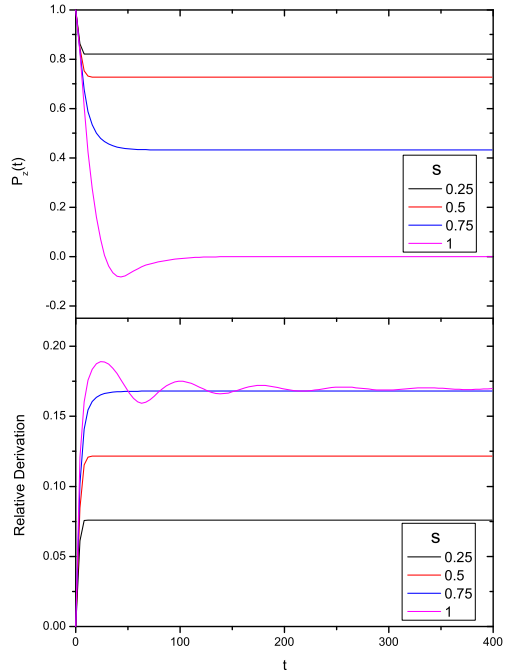}
\caption{Time-dependent population difference $P_z(t)$ (upper panel) and relative deviation $\sigma (t)$ (lower panel) for $\alpha=0.5$, $\omega_c/\Delta=5$, and four values of the exponent $s$: $s=0.25,~0.5,~0.75$, and $1$. The factorized bath initial condition is taken.}
\end{figure}

Thus, the bath initial conditions play an important role in the dynamics of the sub-Ohmic
spin-boson model. In Ref.~\cite{Kast}, the authors studied out-of-equilibrium bath preparations with respect to the initial state of the spin. The polarized bath initial condition corresponds to the case where the bath distribution is fully equilibrated with the initial state of the spin, while the factorized bath initial condition corresponds to the one most displaced from the equilibrium. We note that our results for $\alpha=0.1$ in Fig.~1 and in the upper panel of Fig.~2 are consistent with those in Ref.~\cite{Kast}, where it is found that compared with the polarized initial condition, the factorized one yields decreased oscillation frequency of the dynamics and increased initial loss in population (cf.~the pink lines in Fig.~1 and the upper panel of Fig.~2).

\begin{figure}[t]
\includegraphics[scale=0.35]{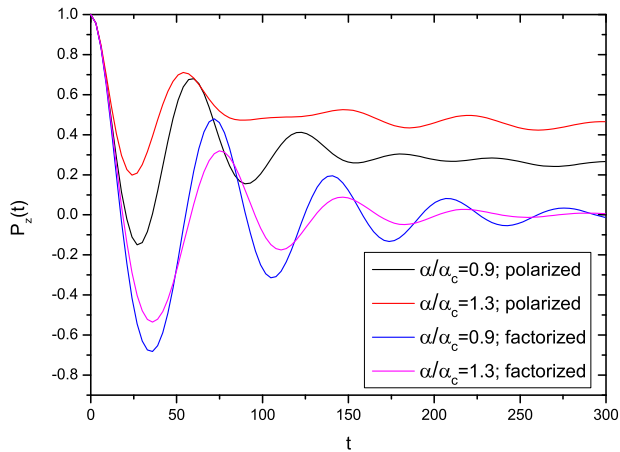}
\caption{Population difference $P_z(t)$ as a function of time for $s=0.25$ ($\alpha_c \approx 0.022$) and $\Delta/\omega_c=0.1$ with polarized and factorized bath initial conditions. }
\end{figure}

An explicit comparison between the polarized and factorized initial conditions is given in Fig.~4, where substantial differences are revealed between the two bath initial conditions for $\alpha=0.9\alpha_c$ and $\alpha=1.3\alpha_c$. Under the factorized bath initial condition, the oscillations occur around zero average in both the localized and delocalized phases, and the population difference $P_z(t)$ is in a delocalized phase even for $\alpha>\alpha_c$, a parameter regime where the state is expected to be localized.
In contrast, the oscillations occur around finite values even in the delocalized phase for the polarized initial condition. Also, the damped constant (or steady state) is an increasing function of $\alpha$.
Similar results are also found by Nalbach and Thorwart using the quasiadiabatic propagator path integral. \cite{Nalbach}.

\begin{figure}[b]
\includegraphics[scale=0.35]{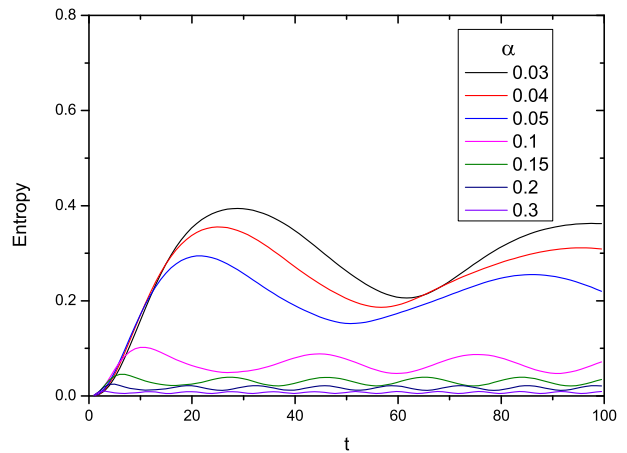}
\caption{ Von Neumann entropy calculated with the polarized bath initial condition for $s=0.25$ ($\alpha_c \approx 0.022$) and $\Delta/\omega_c=0.1$.}
\end{figure}

\begin{figure}[t]
\includegraphics[scale=0.35]{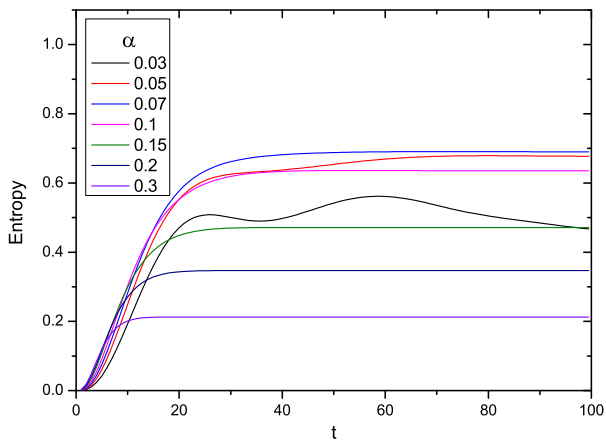}
\caption{Von Neumann entropy calculated under with the factorized bath initial condition for $s=0.25$ ($\alpha_c \approx 0.022$) and $\Delta/\omega_c=0.1$. }
\end{figure}

We also monitor the entanglement between the spin and the bath via the von Neumann entropy. At $t=0$, the systems is in a separable state, so that $S(0)=0$. Eq.~(\ref{entropy}) manifests that the entropy $S$ increases monotonously with $\omega_+$ from $0$ and reaches its maximum at $\omega_+=1/2$, then decreases monotonously to $0$. Figs.~5 and 6 show that there are substantial differences in the time evolution of the entropy under the polarized and factorized bath initial conditions. Under the polarized initial condition, overall the entropy decreases as $\alpha$ increases as shown in Fig.~5. For strong coupling strengths, the entropy eventually vanishes as expected\cite{Costi}. Under the factorized initial condition, the entropy establishes its steady values quickly for various coupling strengths. Interestingly, the steady value is not a monotonous function of $\alpha$, and reaches its maximum at approximately $\alpha\approx0.07$.

\section{Conclusions and Discussions}

Although quite a few numerical approaches have been applied to study the ground state and dynamical properties of the sub-Ohmic spin-boson model, few analytical treatments have been available. Recently, Chin {\it et al}.~\cite{Chin} used an extension of the Silbey-Harris variational wave function to study the ground state of the sub-Ohmic Spin-boson model with $0<s<0.5$, and found that such a trial state generates correct mean-field exponents for the continuous localization-delocalization transition. The asymmetrically displaced-oscillator (ADO) trial state used in Ref.~\cite{Chin} is of the form
 \begin{eqnarray}
|\psi\rangle_{\rm ADO}=&&A|+\rangle\exp[\sum_l (f_lb_l^\dag-h.c.)]|0\rangle_{\rm ph}\nonumber\\
\label{ADO}
&+&B|-\rangle\exp[\sum_l (g_lb_l^\dag-h.c.)]|0\rangle_{\rm ph},
\end{eqnarray}
where the variational parameters $A$, $B$, $f'_ls$ and $g'_ls$ are all real numbers to be determined by the ground-state energy minimization.
It is interesting to note that our ansatz, Eq. (\ref{trial func}), is reduced to Eq.~(\ref{ADO}) if we restrict all variational paremeters to be real, and consequently, the Dirac-Frenkel variation employed in this work is reduced to the conventional variational principle for the ground state used in Ref.~\cite{Chin}. Therefore, we have successfully extended the static trial state of Eq.~(\ref{ADO}) to its dynamical counterpart, which is similar in form to the Davydov $D_1$ ansatz. The foregoing connection also helps map out the validity regime of our ansatz. As pointed out by Nazir {\it et al}., although the ADO state works well for the sub-Ohmic baths with $s<0.5$, it becomes unstable and deviates from the well-established results for the Ohmic case and strong coupling (the cause of which is still under investigation) \cite{ChinPRB}. Thus, it is expected that our ansatz reveals reliable dynamics of the sub-Ohmic spin-boson model in the regime of $s\ll1$, but may lose accuracy for large $s$.

For the sub-Ohmic spin-boson model with $0<s<1/2$, detailed dynamics of the spin-boson model in the strong coupling regime is still surrounded by  contention. The hierarchy of the Davydov ans\"{a}tze of varying sophistication has been known to be competent in handling polaron dynamics in the strong coupling regime \cite{Sun,pssc}. Using a version of the most accurate of the hierarchy, the Davydov ${\rm D}_1$ ansatz, we have carried out a time-dependent variational calculation with regard to the dynamic properties of the sub-Ohmic spin-boson model. It is a simple, yet extremely efficient approach to investigate the dynamics of a quantum dissipative system, such as the population disparity $P_z(t)$ of the sub-Ohmic spin-boson model.
Our results are consistent with those obtained using numerically much more expensive advanced numerical methods, such as the path integral Monte Carlo simulations\cite{Kast,Nalbach} and the ML-MCTDH technique \cite{Wang2}. It is found that the bath initial conditions have considerable influence over the dynamics of this many-body dissipative system. Even in the very strong coupling regime, quantum coherence features still manage to survive under the polarized bath initial condition, while such features are absent under the factorized bath initial condition.
The onset of the incoherent phase occurs at $\alpha=0.1$ for $s=0.25$ under the factorized bath initial condition.
Our findings are consistent with those in Ref.~\cite{Kast}, which first reported the persistence of coherent quantum dynamics at strong dissipation under the polarized bath initial condition.
Furthermore, the Davydov ${\rm D}_1$ ansatz has been employed successfully to study excitation energy transfer across light-harvesting complexes in photosynthesis \cite{Ye}. Our approach may turn out to be a competitive tool to investigate sustained quantum coherence recently discovered to reside in pigment networks even at elevated temperatures \cite{Engel,Calhoun,Collini}.

\section*{Acknowledgements}

Support from the Singapore National Research Foundation through the Competitive Research Programme (CRP)
under Project No.~NRF-CRP5-2009-04 is gratefully acknowledged. The authors thank Q.H.~Chen for useful discussions.

\end{document}